\documentclass[twoside,11pt]{report}

\usepackage[utf8]{inputenc}
\usepackage[T1]{fontenc}
\usepackage{xspace}
\usepackage{times}
\usepackage{calc}
\usepackage{ifthen}
\usepackage{chapterbib}
\usepackage{graphicx}
\usepackage{multirow}
\usepackage{array}
\usepackage{tabularx}
\usepackage{pdfsync}
\usepackage{fancyhdr}
\usepackage{alltt}
\usepackage{hyperref}
\usepackage{rotating}
\usepackage{amsmath}
\usepackage{subfigure}
\usepackage[scaled=0.92]{helvet}

\usepackage{longtable}
\usepackage{array}
\usepackage[table]{xcolor}

\usepackage{url}
\makeatletter
\def\url@leostyle{%
  \@ifundefined{selectfont}{\def\UrlFont{\sf}}{\def\UrlFont{\small\sffamily}}}
\makeatother
\usepackage[english]{babel}

\usepackage{listings}
\usepackage{color}
\usepackage{textcomp}

\definecolor{listinggray}{gray}{0.9}
\definecolor{lbcolor}{rgb}{0.95,0.95,0.95}
\definecolor{lightgray}{rgb}{.9,.9,.9}
\definecolor{darkgray}{rgb}{.4,.4,.4}
\definecolor{purple}{rgb}{0.65, 0.12, 0.82}

\lstdefinelanguage{JavaScript}{
  keywords={typeof, new, true, false, catch, function, return, null, catch, switch, var, if, in, while, do, else, case, break, with, for},
  keywordstyle=\color{blue}\bfseries,
  ndkeywords={class, export, boolean, throw, implements, import, this},
  ndkeywordstyle=\color{darkgray}\bfseries,
  identifierstyle=\color{black},
  sensitive=false,
  comment=[l]{//},
  morecomment=[s]{/*}{*/},
  commentstyle=\color{purple}\ttfamily,
  stringstyle=\color{red}\ttfamily,
  morestring=[b]',
  morestring=[b]"
}

\usepackage{titletoc}
\usepackage{titlesec}
\titleformat{\chapter}[hang]
{\sffamily\huge}
{\chaptertitlename\thechapter.}
{1ex}
{}
[\titlerule]
\titlespacing*{\chapter} {-6pc}{-10pt}{10pt}

\titleformat{\section}
{\Large\sffamily\slshape}
{\thesection}{1em}{}
\titlespacing{\section}
{-4pc}{1.0ex plus .1ex minus .2ex}{1.5ex minus .1ex}

\titleformat{\subsection}
{\large\sffamily\slshape}
{\thesubsection}{1em}{}
\titlespacing{\subsection}
{-2pc}{2.ex plus .1ex minus .2ex}{2.5ex minus .1ex}

\titleformat{\subsubsection}
{\sffamily\slshape}
{\thesubsubsection}{1em}{}
\titlespacing{\subsubsection}
{-1pc}{1.5ex plus .1ex minus .2ex}{1.2ex minus .1ex}

\renewcommand\paragraph[1]{\par\vspace{1.ex}\noindent\textbf{#1}}

\usepackage{amssymb}

\newcommand{\ct}[1]{\textsf{#1}}

\newcommand{\textsfmmented}[1]{}
\newcommand{\doNotShow}[1]{}
\newcommand{\prog}[1]{}

\newcommand{\eg}{\emph{e.g.,}\xspace}
\newcommand{\ie}{\emph{i.e.,}\xspace}

\newboolean{showcomments} \setboolean{showcomments}{true}
\ifthenelse {\boolean{showcomments}} {
  \newcommand{\mynote}[2]{\fbox{\bfseries\sffamily\scriptsize#1}{\small$\blacktriangleright$\textsf{\emph{#2}}$\blacktriangleleft$}}}{
  \newcommand{\mynote}[2]{}}

\graphicspath{{.}{figures/}}

\newcommand{\this}{\textsf{this}\xspace}
\newcommand{\new}{\textsf{new}\xspace}
\newcommand{\javascript}{JavaScript\xspace}
\newcommand{\window}{\textsf{window}\xspace}
\newcommand{\ecma}{ECMAScript\xspace}
\newcommand{\js}{\javascript}

\begin{document}
\title{Semantics and Security Issues in \js\\~\\ \large S\'emantique et probl\`emes de s\'ecurit\'e en \js \\~\\ \large{Deliverable Resilience FUI 12:\\ 7.3.2.1 Failles de s\'ecurit\'e en JavaScript \\/ JavaScript security issues}}
\author{St\'ephane Ducasse, Nicolas Petton, Guillermo Polito, Damien Cassou}
\date{Version 1.2 - December 2012}
\maketitle

~ 
\vfill
\begin{footnotesize}
\setlength{\parindent}{0pt}

Copyright \copyright~2012 by S. Ducasse, N. Petton, G. Polito, D. Cassou.\\[1cm]

The contents of this deliverable are protected under Creative Commons Attribution-Noncommercial-ShareAlike 3.0 Unported license.

\emph{You are free:}
\begin{description}
  \item[to Share] \,---\, to copy, distribute and transmit the work
  \item[to Remix] \,---\, to adapt the work
\end{description}
\emph{Under the following conditions:}
\begin{description}
  \item[Attribution.] You must attribute the work in the manner specified by the author or licensor (but not in any way that suggests that they endorse you or your use of the work).

\item[Noncommercial.] You may not use this work for commercial purposes.

  \item[Share Alike.] If you alter, transform, or build upon this work, you may distribute the resulting work only under the same, similar or a compatible license.
\end{description}
\begin{itemize}
  \item For any reuse or distribution, you must make clear to others the license terms of this work. The best way to do this is with a link to this web page:
  \url{creativecommons.org/licenses/by-sa/3.0/}
  \item Any of the above conditions can be waived if you get permission from the copyright holder.
  \item Nothing in this license impairs or restricts the author's moral rights.
\end{itemize}
\raisebox{-0.25cm}{\includegraphics[width=2cm]{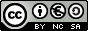}}\quad
\parbox{\textwidth}{
	Your fair dealing and other rights are in no way affected by the above.
	This is a human-readable summary of the Legal Code (the full license):\\
	\url{http://creativecommons.org/licenses/by-nc-sa/3.0/legalcode}}\\[1cm]
\end{footnotesize}
\vfill

\newpage

\paragraph{Deliverable: 7.3.2.1 Failles de s\'ecurit\'e en Javascript}
\paragraph{Title: Semantics and Security Issues in \js}
\paragraph{Titre: Sémantique et problèmes de sécurité en \js}
\paragraph{Version: 1.2}
\paragraph{Authors:   St\'ephane Ducasse, Nicolas Petton, Guillermo Polito, Damien Cassou}

\paragraph{Description de la t\^ache 7.3.2.1 / Task 7.3.2.1  Description:}

Les environnements d'ex\'ecution Javascript comportent des failles de s\'ecurit\'e li\'ees \`a la s\'emantique d'ECMAScript ainsi qu'\`a la pr\'esence de certains constructeurs. D'autres failles sont li\'ees \`a la mise en oeuvre pratique d'ECMAScript par les navigateurs. L'objectif de cette t\^ache est d'identifier et classer les types de failles de s\'ecurit\'e dans le monde Javascript. On consultera notamment les travaux de Miller autour de Caja et ainsi que les approches propos\'ees par Seaside un framework pour le d\'eveloppement d'applications s\'ecuris\'ees en Smalltalk. L'objectif de la t\^ache est de formaliser les m\'ecanismes de s\'ecurit\'e en Javascript de fa\c con \`a augmenter la connaissance des partenaires et des utilisateurs de Javascript sur ce sujet.


\tableofcontents

\newpage

\begin{abstract}
  There is a plethora of research articles describing the deep
  semantics of \js. Nevertheless, such articles are often difficult to
  grasp for readers not familiar with formal semantics. In this
  report, we propose a digest of the semantics of \js centered around
  security concerns.  This document proposes an overview of the \js
  language and the misleading semantic points in its design. The first
  part of the document describes the main characteristics of the
  language itself. The second part presents how those characteristics
  can lead to problems. It finishes by showing some coding patterns to
  avoid certain traps and presents some \ecma 5 new features.
\end{abstract}

\newpage

\chapter{Web architecture basics}

In this chapter we present some basics principles of common web
architecture and in which \js applications are deployed.

\section{Client-Server basics}
The web is based on the client-server architectural pattern:

\begin{itemize}
\item the client role is played by a web browser with limited
  resources and technologies -- often HTML, CSS and \js. Its usual
  main responsibilities are user interface and interaction as well as
  data validation.
\item the server role is fulfilled by a program implemented in a wide
  variety of technologies, with a controlled set of resources. Its
  usual responsibilities are to serve pages to web-browsers, enforce
  business rules, and persist and validate data.
\end{itemize}

\begin{figure}
\centering
\includegraphics[width=10cm]{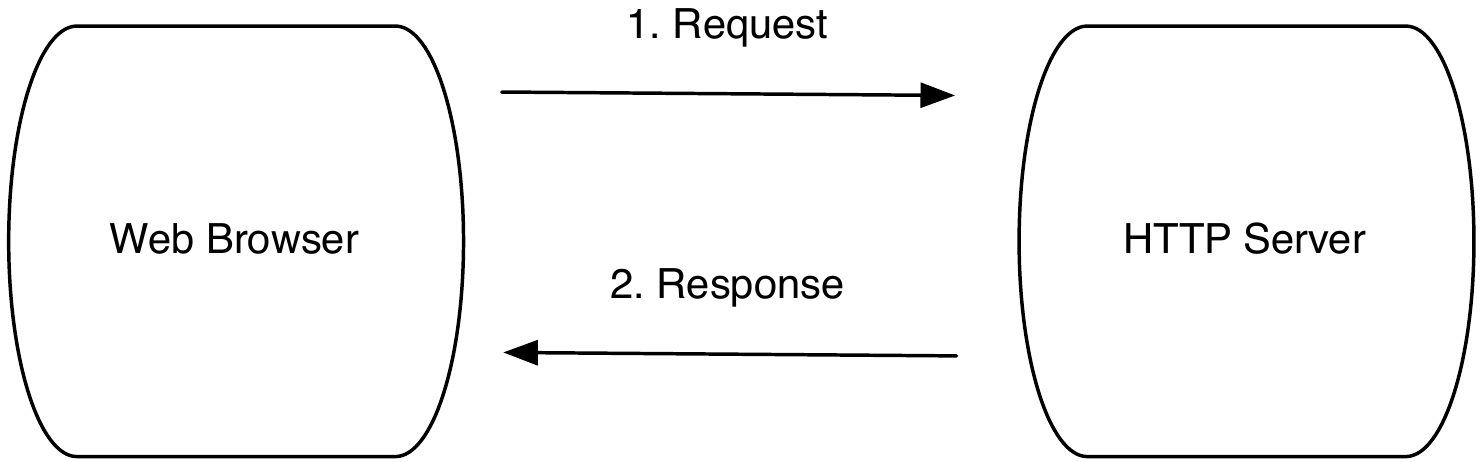}
\caption{An HTTP request and response between a web browser and a web server\label{fig:httpRequest}}
\end{figure}

The client and server processes communicate though the HTTP protocol
\cite{Mogu02a}: the client makes HTTP requests and, for each, the
server answers with an HTTP response, as illustrated in
\autoref{fig:httpRequest}. When the response arrives, the web browser
normally discards the current content and loads the new one, causing a
full \emph{refresh} of the user interface. Since HTTP is a stateless
protocol -- requests are independent from each others -- several
mechanisms were built to maintain state between a client and a server
(\eg{} cookies, see \autoref{sec:cookies}).

As computers running \emph{web browsers} become more and more
powerful, there is a tendency to move responsibilities from the server
to the client (\eg{} page creation and business rules
enforcement). These new responsibilites are mostly implemented in \js
and \js-based languages.

\section{Web Browser window insfrastructure}
\label{sec:browser}

A web page is composed of several html widgets including frames and
iframes, which insert an existing web page into the current
one. \emph{Frames} and \emph{iFrames} are considered web pages by
themselves, therefore having the ability to contain other ones,
following the composite design pattern \cite{Gamm95a}.

While loading a web page, a web browser generates an internal
structure of \js objects. Each web page is represented as a \window
object referencing its parent and children through the \ct{parent} and
\ct{frames} properties.

Each web page also references a document object, which is the root of
the \emph{DOM} -- Document Object Model -- representation of the web
page tags.

\begin{figure}
\centering
\includegraphics[width=10cm]{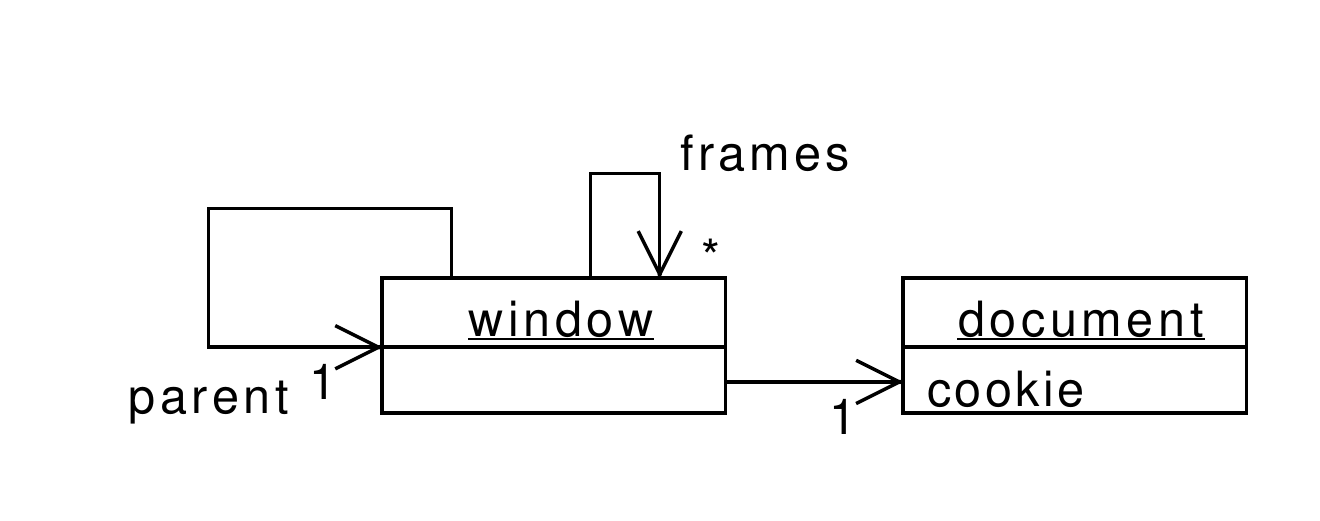}
\caption{The \js model of a web page\label{fig:documentStructure}}
\end{figure}

\paragraph{Web browser \js implementation differences}

As each web browser has its own implementation of the \emph{DOM} and
\js, one of the hardest tasks of \js development is to build
browser-agnostic applications. There exist several libraries to reduce
the difficulty of this task such as
\emph{extJS}\footnote{http://www.sencha.com/products/extjs/},\emph{jQuery}\footnote{http://jquery.com/}
and \emph{scriptaculous}.\footnote{http://script.aculo.us/}

\section{Cookies}
\label{sec:cookies}

A cookie is a string that is exchanged between the client and the
server \cite{Mill01c}. Cookies are normally stored on the client
machines to let web application state survive after the client process
is stopped (\ie the browser or the browsing tab is closed).

Cookies are key-value pairs associated with the equals character '='
and separated from each other with a semi colon (\eg,
\emph{"name=john;expires=2013-01-15T21:47:38Z"}). The expiration date
tells the browser when to delete it. A cookie can contain other
properties like \ct{domain} and \ct{path}, that tells the browser in
which requests it should be exchanged, which by default are the
current domain and path.

Cookies can be changed from \js code accessing the \ct{cookie}
property of the current \ct{document} object.  Cookies can contain
sensible data (\eg emails and passwords) and must be studied for a
security analysis.

\section{AJAX}
\label{sec:ajax}
\emph{AJAX} stands for Asynchronous \js and XML: it is a mechanism to
send XMLHttpRequest requests to a web server.  XMLHttpRequest is used
to retrieve data from a URL without reloading the complete web page
\cite{Mesb07a}. The request is sent by the client through a \js API
and handled asynchronously. Formerly, AJAX was used to exchange
information in XML format, but XMLHttpRequest can be used to transfer
any kind of data. Nowadays, one of the most used formats is
\emph{JSON},\footnote{\url{http://www.json.org}} which stands for \js
Object Notation.

Illustrated in \autoref{lst:ajax}, the "test.json" service is
requested to the server via AJAX, and a function is used to determine
how the response will be handled at the time it arrives.

\begin{lstlisting}[caption=AJAX request performed with JQuery Library, label={lst:ajax}]
$.ajax({
    url: "http://www.webcompany.com/api/test.json",
}).done(
    // Function is evaluated  after the json service responds:
function() {
    // adds a 'class' attribute to current HTML node
    $(this).addClass("done");
});
\end{lstlisting}

AJAX is becoming more and more popular as the client machines become
more powerful and can perform heavier processing tasks.

\chapter{\js object model in a nutshell}

In this chapter we present key aspects of \ecma 3 and 5\footnote{\ecma
  is the standard \js is based on.} to provide a common context. This
document has been written in the context of the Resilience project.
This study is focused on the semantic aspects of \js in the context of
securing untrusted third party \js code in a web browser. In this
context, one of the key aspects is to forbid the access to the \window
object (the global object of the web browser frame, see
\ref{sec:window}). Accessing the \window object of the frame would
lead to possible page-data leak (\window is the root of the DOM of the
page), sensitive information leak (from the \window object private
information in cookies can be accessed), and even user actions
sniffing (having access to DOM elements means that one can add
listeners to DOM events\footnote{Adding event listeners to DOM objects
  allows an attacker to catch user interactions and inputs.}). In the
next deliverable we will describe the state of the art of \js
sandboxing. One of the different aspects in \js sandboxing is to
securely forbid access to potentially sensitive objects, including
\window, the root object.

\section{Objects: the basic building block of \js}
\label{sec:objects}

\js is a loosely-typed dynamic programming language where everything
is an object. Each object contains a set of properties (also named
slots) that represent data (other objects) and operations (function
objects).  These properties are always public and can be accessed and
modified using the dot or squared-bracket notation:

\begin{lstlisting}[caption=Property access]
// We create a new empty object
var person = new Object();

// Write a property
person.age = 5;
person['age'] = 5; // equivalent

// Read a property and store it into some variable
var theAge = person.age;
var theAge = person['age']; // equivalent
\end{lstlisting}

In this script we create a new object that we assign to the variable
\ct{person}. Then the expression \ct{person.age = 5} adds a property
to the newly created object and stores the value 5 into it. Note that
objects are hash tables. The expression \ct{\{a: 0\}} thus creates an
object with one property named \ct{a} with value 0. Because of this,
properties can be accessed and set using both the \ct{person['age']}
and \ct{person.age} notations.

\section{Property access}

ECMAScript's property lookup is done at runtime (see
\autoref{lst:runtime_property}) and never throws errors.  Example
\ref{lst:runtime_property} defines an object \ct{object} containing
two properties \ct{a} and \ct{b} with values 10 and 5 respectively.
The function \ct{get} is then defined. This function takes as argument
a property's name and returns the corresponding property's value of
the object. In the last line of Example \ref{lst:runtime_property},
the function \ct{get} is called twice to access the values of the
properties \ct{a} and \ct{b}.

\begin{lstlisting}[caption=Property lookup at runtime, label={lst:runtime_property}]
var object = {a : 10, b: 5};
var get = function(property) {
  return object[property]
};
get("a") + get("b"); // answers 15
\end{lstlisting}

If a property does not exist, the \ct{undefined} object is returned,
as shown in \autoref{lst:inexisting_property}.

\begin{lstlisting}[caption=Property access for nonexistent properties, label={lst:inexisting_property}]
var object = {a: 1, b: 2};
object.c // answers 'undefined'
\end{lstlisting}

Example \ref{lst:updating_properties} shows how to update, create and
delete properties.  The first instruction sets 10 to the property
\ct{a} of object \ct{\{a: 0\}}. Therefore the expression returns an
object with the property \ct{a} with 10 as a value.

The second instruction shows that if a property does not exist then it will be created automatically. Hence,
\ct{\{a: 0\}[b]=10} returns an object with two properties \ct{a} and \ct{b}.

Finally, the third instruction removes the property \ct{b} of the object using the \ct{delete} keyword.

\begin{lstlisting}[caption={Updating, creating and deleting properties}, label={lst:updating_properties}]
// Updating a property
{a: 0}[a] = 10; // answers {a: 10}

// Creating a new property using the same syntax
{a: 0}[b] = 10; // answers {a: 0, b: 10}

// Deleting a property
delete {a: 0, b: 5}[b]; // answers {a: 0}
\end{lstlisting}

\section{Functions: behavioral units}
\label{sec:functions}

Functions are the behavioral units of \js. A function performs a side
effect (\eg alters an object) and/or returns a value.

Functions can be defined via a function declaration or a function
expression. A function declaration defines a function at compile time,
as seen in \autoref{lst:functionDeclaration}. A function expression
creates a new function at runtime. Function expressions are also
called anonymous functions, since they are not associated with any
name. The creation of a function from a function expression is
illustrated in \autoref{lst:functionExpression}.

\begin{lstlisting}[caption=A function declaration has a name, label={lst:functionDeclaration}]
function sum(a, b){
    return a + b;
}
\end{lstlisting}

\begin{lstlisting}[caption={An anonymous function that is assigned to a variable}, label={lst:functionExpression}]
var sum = function (a, b){
    return a + b;
}
\end{lstlisting}

A function can be called by passing arguments between parenthesis as
shown in \autoref{lst:functionCall}.

\begin{lstlisting}[caption={Calling a function}, label={lst:functionCall}]
sum(1, 2);
someFunctionWithNoArguments();
\end{lstlisting}

Functions are important because they constitute the basic building
block of code execution in a \js program.

\section{Programmatic function evaluation}

The built-in \ct{call()} and \ct{apply()} functions are two methods of
the \ct{Func\-tion} object.\footnote{In \js, the words 'method' and
  'function' are equivalent. The former is particularly used when the
  function is owned by an object as one of its properties.} These
functions are thus called on other functions and execute them. These
functions both take the receiver of the function to execute as first
parameter and arguments of the function to execute as the other
parameters. When using these methods, the pseudo-variable \this in the
function to execute is bound to the first parameter of \ct{call()} and
\ct{apply()}. \ct{call()} receives the arguments of the function to
execute as comma-separated values whereas \ct{apply()} receives an
array (see \autoref{lst:apply_call}).

\begin{lstlisting}[caption=The \ct{apply()} and \ct{call()} methods, label={lst:apply_call}]
function someGlobalFunction(value1, value2){
	this.value1 = value1;
	this.value2 = value2;
}

// The regular invocation binds 'this' to the global object as
// we will see later on

someGlobalFunction(5,6);
window.value1       // answers 5

var someObject = new Object();
someGlobalFunction.call(someObject, 5, 6);
someObject.value1 // answers 5

someGlobalFunction.apply(someObject, [5, 6]); // equivalent

\end{lstlisting}

\paragraph{Turning text into function at runtime.} Using the
\ct{eval()} built-in function, it is possible to evaluate a string at
runtime.  \ct{eval()} makes static analysis of programs difficult
\cite{Rich11a}: at runtime a string can be composed and evaluated
without restriction in the global environment and scope.

\begin{lstlisting}[caption=Evaluating code from a string]
var a = 2;
eval('a++');
a // answers 3
\end{lstlisting}

\section{Object methods}

In \js, methods are just functions attached to an object as can be
seen in \autoref{lst:adding_methods}.

\begin{lstlisting}[caption={Adding methods to an object}, label={lst:adding_methods}]
var person = new Object();
person.name = 'John';
person.surname = 'Foo';
person.getFullName = function (){
    return this.name + ' ' + this.surname;
}
\end{lstlisting}

In a method, a developer can refer to the owner object using the \this
keyword.  We will see later that \ct{this} has a different semantics
than in Java or Smalltalk since depending on how the function is
accessed \this can be bound to the object owner of the property or any
other object (see \autoref{sec:this}).

\section{Higher-order functions}

\emph{Functions} in \js, as first class citizens, can be passed as
parameters to functions and returned from
functions. \emph{Higher-order functions} are most commonly seen when
implementing filtering, selection and sorting algorithms to make them
independent of their content. In \autoref{lst:high_order}, we define a
new property inherited by all arrays that is a function to filter
elements of the receiver array according to a given criteria. Then, we
define a function \ct{isEven} that will play the role of the criteria
and an array of numbers. The last statement calls the new \ct{filter}
property function on \ct{array} with the \ct{isEven} criteria.

\begin{lstlisting}[caption=Extending Arrays with a higher-order filter function, label={lst:high_order}]
Array.prototype.filter = function (criteria){
    newArray = new Array();
    for (var i = 0; i < this.length; i++) {
        // we keep the elements of the array that respects
        // a certain criteria
        if (criteria(this[i]))
            newArray.push(this[i]);
    }
    return newArray;
}

var isEven = function(elem) { return (elem % 2) == 0; };
var array = new Array(9, 58, 42, 12, 1001, 1000);
array.filter(isEven); // answers [58, 42, 12, 1000]
\end{lstlisting}

\section{Object constructors}

\emph{Constructors} are used to structure object creation in
\ecma3\cite{Maff08a, Guha10a}. A constructor is a standard function
object whose name is by convention capitalized to indicate to the
programmer that the function must not be directly called. The \ct{new}
keyword is used to invoke a function as a constructor. Using the
\ct{new} keyword instantiates an object and executes the constructor
on that object, binding the pseudo-variable \this to the newly created
object.

\begin{lstlisting}[caption=The \ct{new} keyword,label={lst:new_keyword}]
var Animal = function (name) {
    this.name = name;
    this.describe = function() {
        return this.name + ', an animal';
    }
};

//Invoking the constructor with the 'new' keyword
var animal = new Animal("pilou");

animal.name;       // 'pilou'
animal.describe()  // 'pilou, an animal'
\end{lstlisting}

\section{Object core properties}

There are three properties that are key to the execution of any \js application:

\newcommand{\proto}{\_\_proto\_\_}

\begin{figure}
  \centering
  \includegraphics[width=.8\textwidth]{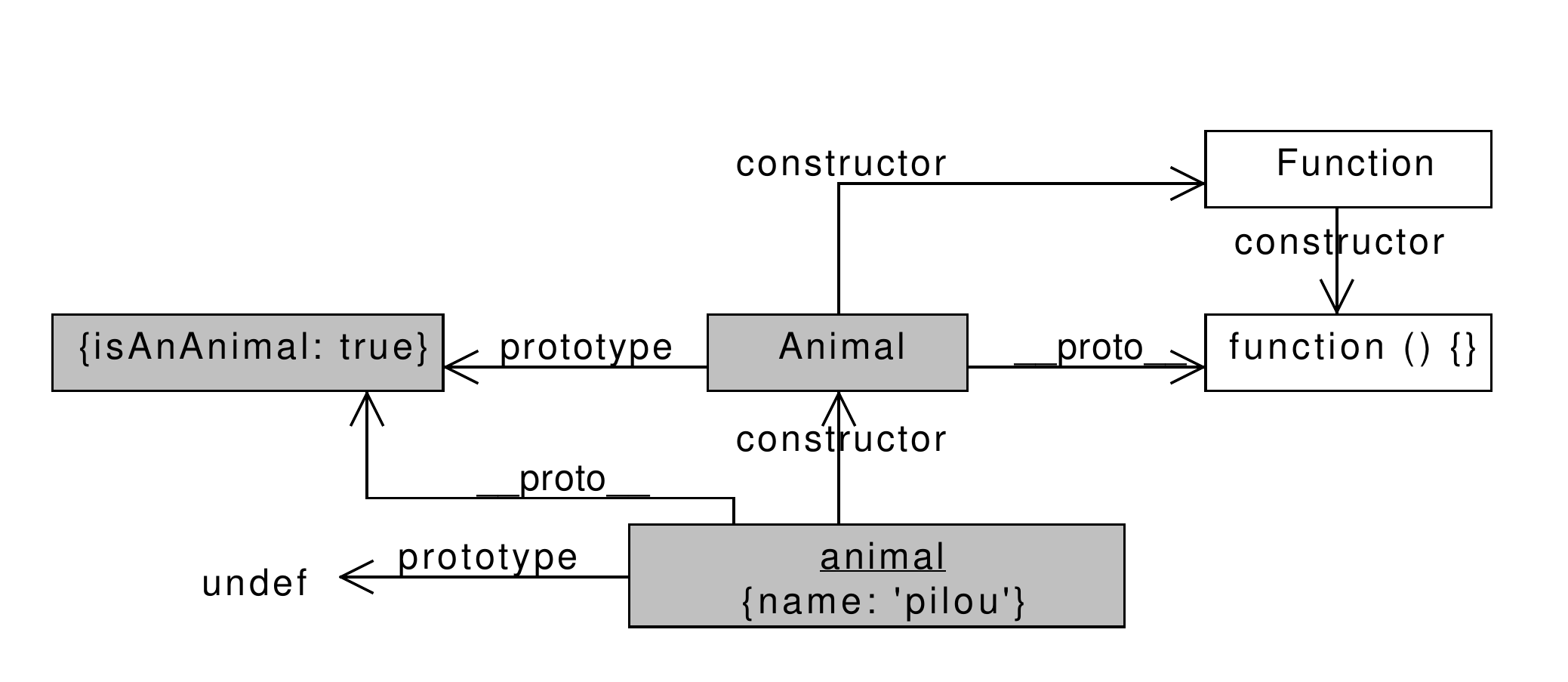}
  \caption{The 3 core properties of objects (\ct{constructor}, \ct{prototype}, \ct{\proto}). Boxes in gray represent objects created by the code of \autoref{lst:core-properties-simple} whereas boxes in white represent objects provided by \js.}
  \label{fig:core-properties-simple}
\end{figure}

\begin{itemize}
\item \ct{constructor} is a property that contains the reference to
  the function that was used to create the object using the \new
  keyword. In \autoref{fig:core-properties-simple}, the \ct{animal}
  object has a property \ct{constructor} containing a reference to the
  \ct{Animal} constructor that was used to create \ct{animal}.
\item \ct{prototype} is a property that is used by constructors to
  initialize the inheritance chain (modeled by the \ct{\proto}
  property) of objects they create. This property only makes sense on
  function objects that are used as constructors. In
  \autoref{fig:core-properties-simple}, the \ct{Animal} constructor
  has a property \ct{prototype} containing a reference to an object
  (with an \ct{isAnAnimal} property) that will be the parent of all
  objects \ct{Animal} creates.
\item \ct{\proto} is a property that contains a reference to an object
  (the parent) where properties will be looked up when not present in
  the receiver. When an object is created, the value of its
  \ct{\proto} property is initialized to the \ct{prototype} property
  of the constructor function used to create this object. In
  \autoref{fig:core-properties-simple}, the \ct{animal} object has as
  parent the object whose only property is \ct{isAnAnimal}: this
  object is the prototype of the \ct{Animal} constructor.
  \emph{Attention:} this property is not always visible from a
  developer's point of view depending on the \js implementation: you
  should avoid manipulating \ct{\proto} if you want to write portable
  code.
\end{itemize}

\autoref{lst:core-properties-simple} demonstrates that the \ct{\proto}
property is initialized to the constructor's \ct{prototype} property
value. First, a function is defined and assigned to a new variable
\ct{Animal}. The prototype of new functions is always an empty object.
In this case, we add the property \ct{isAnAnimal} to this empty
object. All objects created from the \ct{Animal} function will then
inherit the \ct{isAnAnimal} property. The variable \ct{animal} is then
set to a new \ct{Animal}. The rest of the statements describe the
relationship between \ct{animal} and its constructor \ct{Animal}.

\begin{lstlisting}[caption={Defining a constructor so that \autoref{fig:core-properties-simple} can show the core properties involved}, label={lst:core-properties-simple}]
var Animal = function (name) { this.name = name; };
Animal.prototype.isAnAnimal = true;

var animal = new Animal("pilou");

animal.constructor == Animal; // answers true
animal.__proto__ == animal.constructor.prototype;   // answers true
animal.isAnAnimal;  // answers true
Animal.isAnAnimal; // answers false, 'isAnAnimal' is not a property of
                    // the constructor but of objects it constructs
\end{lstlisting}

\chapter{Key subtle \js points}

In the following we present the subtle aspects of \js that have an
impact on security such as unbound variables \cite{Maff08a, Flan06c}.

\section{Variables and scope}

All objects inside a web browser frame share the same environment
without any restriction. This is a security problem because it allows
dynamic modifications of the program and full access to all the
objects defined inside this environment.

To declare a variable local in a particular scope such as a function
scope the \ct{var} keyword is needed, as illustrated in
\autoref{lst:localVariable}. Here the variable
\ct{my\-Local\-Variable} is local and only accessible from within the
body of the function \ct{myFunction}.

\begin{lstlisting}[caption=declaring a variable local to the function, label={lst:localVariable}]
function myFunction(arg) {
	var myLocalVariable = arg + 5;
	return myLocalVariable;
}
\end{lstlisting}



Not using \ct{var} causes our program to create a global variable (see \autoref{lst:globalVariable}) \cite{Maff08a, Flan06c, Guha10a}.
\begin{lstlisting}[caption=using a global variable, label={lst:globalVariable}]
(function () { globalVar = 'setting global'; })()

window.globalVar // answers 'setting global'
\end{lstlisting}

In this example, \ct{globalVar} becomes a property of the global
environment \window.

\section{The \window object}
\label{sec:window}

The \window object represents the \window of the current HTML
document. Each web browser's tab, frame, or window has its own \window
object.

The \window object is the global object of the \js environment. Each
function that is not attached to another object is attached to
\window, as illustrated in \autoref{lst:windowglobal}. This behavior
combined with the dynamic binding of the \this pseudo-variable (see
\ref{sec:this}) can have deep side effects on security of your
applications.

\begin{lstlisting}[caption={A function not attached to an object is attached to \window}, label={lst:windowglobal}]
window.ping // answers 'undefined'

var ping = function (string) {
  return string;
};

window.ping; // answers the ping function
window.ping('foo'); // answers 'foo'
\end{lstlisting}

The example starts by checking that the \window object has no property
named \ct{ping}. Then we define a new function at the top level, which
is automatically added to the \window object. Accessing
\ct{window.ping} returns the \ct{ping} function. We check that we can
effectively execute it.

\section{\this: an overly dynamic pseudo-variable}
\label{sec:this}

While at first sight, \this may look like the pseudo-variable in Java,
C\# and Smalltalk (named \ct{self} in this case), in \js \this has a
special semantics.  \this is a dynamically bound builtin
pseudo-variable of \js.  When used inside a function scope, the value
of the pseudo-variable \this is determined by the syntactic shape of
the function invocation \cite{Maff08a, Guha10a}. By syntactic shape,
we mean that \this is bound to a different object depending on the way
the invocation expression is written.

In \autoref{lst:function_owner}, while the same function is shared
between two different objects, depending on the function invocation,
\this is bound to a different object.

\begin{lstlisting}[caption=Function invoking, label={lst:function_owner}]
var o = new Object();
// A new function is attached to 'o' on property 'f'
o.f = function() {return this;};

// f is invoked from o
o.f(); // answers o, so 'this' was bound to 'o'

var o2 = new Object();

// o2.f and o.f point to the same function object
o2.f = o.f;

// f is invoked from o2
o2.f(); // answers o2, so 'this' was bound to 'o2'
\end{lstlisting}

The behavior described by the previous example looks natural and
similar to the one of Java and C\# where \this is bound to the
receiver. However, the semantics behind \ct{this} is more
complex. Thus, depending on the way the function is invoked, \ct{this}
may not refer to the object which defines the function but to the
global object \window.  As explained in \autoref{sec:window}, every
object not attached to another object is implicitly attached to the
global object \window.

The following example illustrates this behavior. First, a simple
object, named \ct{o}, with a method \ct{yourself} is created that
simply returns \ct{this}. When the expression \ct{o.yourself} is
executed, \ct{o} is returned as expected.  Then the variable
\ct{yourselfFunction} is defined in the context of the global
environment and its value is set to the method \ct{yourself}. Now when
\ct{yourselfFunction} is executed, the global object \window is
returned instead of the object \ct{o}, which defines the method.

We see that assigning a pointer to a function and invoking this
function via this pointer changes what \this is bound to inside the
function.

\begin{lstlisting}[caption={\this and \window interplay}, label={lst:thiswindow1}]
// Creates a new object with a 'yourself' method
var o = {
  yourself: function() { return this; }
};

o.yourself() // answers o

//We attach o.yourself to window
var yourselfFunction = o.yourself;

yourselfFunction() // answers the 'window' object
\end{lstlisting}

As can be seen in \autoref{lst:thiswindow1} taken from \cite{Guha10a},
one of the dangerous side effects of the \this semantics is the
ability to retrieve the \window object from a function.

In this example, an object, named \ct{obj}, is created with a property
\ct{x} and a method named \ct{setX:} which mutates \ct{x}. The return
value of the expression \ct{window.x} shows that \window does not have
a property \ct{x}. Then the expression \ct{obj.setX(10)} sets the
value of the property \ct{x} of \ct{obj} to \ct{10}.  Then the
variable named \ct{f} points to the mutator method of object \ct{obj}.

Executing the mutator via the reference through \ct{f} with \ct{90} as
a value will add a variable \ct{x} to \window and assign it
\ct{90}. Here the syntax of the method invocation binds \this to the
global object \window instead of \ct{obj}. The value returned from the
expression \ct{obj.x} shows that the property \ct{x} did not change
and is still holding \ct{10}. Finally, the value returned from the
expression \ct{window.x} shows that the object \window got a new
property \ct{x} holding the value \ct{90}.

\begin{lstlisting}[caption=\this and \window, label={lst:thiswindow2}]
var obj = {
  "x" : 0,
  "setX": function(val) { this.x = val }
};

window.x     // answers 'undefined'
obj.setX(10);
obj.x         // answers 10
var f = obj.setX;
f(90);
obj.x // answers 10 (obj.x was not updated)
window.x // answers 90 (window.x was created)
\end{lstlisting}

This section has shown that \this is dynamically bound and that its
binding depends on the syntactic expression from which a function is
invoked. In particular, imprecise use of \this may lead to a security
breach granting access to the top-level object \window (and thus any
object of the web browser frame).

\section{Object constructors misuse}
\label{sec:constructors}

\paragraph{Constructors used without new.} When invoked without the
\ct{new} keyword, the function gets evaluated in the global
context. \this is then bound to the \window object as seen in
\autoref{lst:notnew_keyword}.

\begin{lstlisting}[caption=Not using the \ct{new} keyword,label={lst:notnew_keyword}]
var Person = function (name, surname, age) {
  this.name = name;
  this.surname = surname;
  this.age = age;
};

// Invoking the constructor as a simple function
var person = Person('John', 'Foo', 27);
person // answers 'undefined'
person.age // raises an error
window.surname // answers 'Foo'
\end{lstlisting}

Note that in \autoref{lst:notnew_keyword}, \ct{person} is undefined
since the constructor does not return the object. In addition \window
gets an additional \ct{surname} property as shown by the last
statement of the example.

Objects created using the same constructor will not share functions
and data set in the constructor. Each object will have its own copy of
the initialization values. Sharing can be achieved using the prototype
as will be shown in \autoref{sec:prototypes}.

\paragraph{Constructors returning objects.}

\autoref{lst:constructor_returning_object} shows that values
returned by functions invoked as constructors (with the \ct{new}
operator) are ignored by the compiler.

\begin{lstlisting}[caption={Returning a primitive type from a constructor function},label={lst:constructor_returning_object}]
var Dog = function () {
  this.name = 'milou';
  return 3; // this statement is ignored by the compiler
}

var dog = new Dog();
dog; // answers {name: 'milou'} as expected
\end{lstlisting}

\autoref{lst:constructor_returning_wrongObject} shows that when the
returned object of a constructor function is not a primitive one, the
constructor actually returns it. This behavior is unspecified by \ecma
5 and leads to misleading results. Constructor functions should never
explicitly return anything. The \new keyword takes care of returning
the newly created object.

\begin{lstlisting}[caption={Returning a non-primitive type from a constructor function}, label={lst:constructor_returning_wrongObject}]
var Dog = function () {
  this.name = 'milou';
  return {name: 'tintin'}; // this statement is not ignored
}

var dog = new Dog();
dog; // unexpectedly answers {name: 'tintin'}
\end{lstlisting}

\section{Lexical closures and unbound variables}
\label{sec:closures}
\js functions are lexical closures \cite{Maff08a, Flan06c,
  Guha10a}. Each lexical environment depends on its outer context. The
closure scope grants access to the function arguments (accessed by
value), as well as all variables accessible from the outer context
scope, including all global variables.

In the \autoref{lst:function_scope}, we show how a function has access
to its outer scope's variables. \ct{outerFunction} is a function which
returns another function whose role is to increment a private variable
\ct{localToOuterFunction} and set the value to an object's property
\ct{someProperty}. We can see that \ct{innerFunction} has access to
the \ct{localToOuterFunction} variable defined in
\ct{outer\-Func\-tion}. We can also see that the two functions
returned by the two calls to \ct{outer\-Func\-tion} have access to two
different copies of \ct{localToOuterFunction}.

\begin{lstlisting}[caption=A function and its outer scope,label={lst:function_scope}]
function outerFunction(obj){
    var localToOuterFunction = 0;
    var innerFunction = function(){
        localToOuterFunction++;
        obj.someProperty = localToOuterFunction;
    }
    return innerFunction;
}
o = new Object();
returnedFunction = outerFunction(o);
returnedFunction();
returnedFunction();
o.someProperty // answers 2

o2 = new Object();
returnedFunction = outerFunction(o2);
returnedFunction();
o2.someProperty // answers 1
o.someProperty // answers 2
\end{lstlisting}

A naive use of closures may lead to issues as described in
\autoref{lst:bad_closures} where all handlers will unexpectedly always
return the value \ct{10}.

\begin{lstlisting}[caption=Variable scopes in closures \#1,label={lst:bad_closures}]
var handlers = [];
for(var i=0; i < 10; i++) {
  handlers[i] = function() { return i; };
};
handlers[3](); // answers 10
\end{lstlisting}

In the \ct{for} loop we iterate over \ct{i} from \ct{0} to \ct{10}. In
each iteration of the loop a closure is created and stored into the
\ct{handlers} array. When one of these closures is evaluated (the
fourth one here), the value returned by the closure is the current
value of \ct{i}, not the value of \ct{i} at the time of the closure
creation.

The \autoref{lst:good_closures} illustrates how to use closures to
solve the issue described before.  This example is the same as the
previous one, only surrounding the closure creation with an additional
closure capturing the value of \ct{i} inside the \ct{for} loop. When
one of the closures is evaluated, the expected number is returned.

\begin{lstlisting}[caption=Variable scopes in closure \#2,label={lst:good_closures}]
var handlers = [];
for(var i=0; i < 10; i++) {
  (function (j) {
     handlers[j] = function() { return j; };
   })(i);
};
handlers[3](); // answers 3
\end{lstlisting}

\section{The \ct{with} statement}
\label{sec:withStatement}

As described in \ecma 3, the \js \ct{with} statement adds all
properties of an object to the scope of the \ct{with} statement as
shown in \autoref{lst:withMix}.

\begin{lstlisting}[caption=Mixing scopes, label={lst:withMix}]
var someGlobal = 'pilou';
var obj = new Object();
obj.propertyA = 1;

with (obj) {
    someGlobal = propertyA;
};

someGlobal; // answers 1
\end{lstlisting}

The scope of the \ct{with} statement includes all the variables of its
outer scope (including global variables) and the object properties,
overriding outer scope variables as shown in
\autoref{lst:withOverride}. In this example, inside the \ct{with}
statement, there is potentially two targets for the \ct{propertyA}
name: this name could be referring to either the global variable (with
value \ct{'property'}) or to the property of \ct{obj} (with value
\ct{1}). When using \ct{with}, properties of the object passed as
parameter to \ct{with} always take precedence.

\begin{lstlisting}[caption=Overriding outer scope variables, label={lst:withOverride}]
var propertyA = 'property';
var someGlobal = 'pilou';
var obj = new Object();
obj.propertyA = 1;

with (obj) {
    someGlobal = propertyA; // 'propertyA' is the property of obj
};

someGlobal; // answers 1
\end{lstlisting}

Using \ct{with} is not recommended and is even forbidden in \ecma 5
strict mode. The recommended alternative is to assign the object's
wanted properties to a temporary variable.

The dynamic nature of \js combined with scope mixture lowers the
predictability of \js programs.

\section{Lifted Variable Definitions}


Local variables are automatically lifted to the top of the function in
which they appear.  For example in the following function \ct{foo()},
the return statement has access to the variable \ct{x} that was
defined inside the if branch.

\begin{lstlisting}
function foo() {
    if (true) {
        var x = 10;
    }
    return x;
}

foo(); // answers 10
\end{lstlisting}

This happens because Javascript automatically rewrite the previous
code to something like the following:

\begin{lstlisting}
function foo() {
    var x;
    if (true) {
        x = 10;
    }
    return x;
}

foo(); // answers 10
\end{lstlisting}

Such behavior can lead to unintuitive results as demonstrated in the
following example.

\begin{lstlisting}
function foo(x) {
    return function() {
        var x = x;
        return x;
    }
}

foo(200)(); // answers undefined
\end{lstlisting}

The function returned by the function \ct{foo} does not return the
value passed to \ct{foo} but \ct{undefined} (the value of unaffected
variables). In this last example we might expect the \ct{x} on the
right-hand side of \ct{var x = x} to reference the argument \ct{x} of
\ct{foo}. Nevertheless, due to lifting, all bound occurrences of
\ct{x} in the nested function refer to the local variable \ct{x} as is
made clear in the following rewrite:

\begin{lstlisting}
function foo(x) {
    return function() {
        var x; // JavaScript splits "var x = x" in two statements
        x = x;
        return x;
    }
}

foo(200)(); // answers undefined
\end{lstlisting}

The expression \ct{var x = x} reads and writes back the initial value
of x (\ie undefined). \ct{var x = x} works as \ct{var x; x = x;} hence
the right-hand \ct{x} of the assignment refer to the local variable
\ct{x}.

\section{Type coercion}

\js performs type coercion (implicit type conversion) over objects on,
for example, comparisons and \ct{if} statements checks. The automatic
type coercions performed by \js are well known causes of bugs which
lower the robustness of \js applications.

The basic rules about type coercion are \cite{ECMA11a}:

\paragraph{Boolean coercion.} When expecting a boolean value, as in
the \ct{if} statements, \js transforms the result of our expression
automatically into a boolean value. Values equivalent to \ct{false}
are \ct{null}, \ct{un\-defined}, \ct{false}, \ct{+0}, \ct{-0},
\ct{NaN} and the empty string. The rest of the \js objects are coerced
to true.

\begin{lstlisting}
var falsyValue = false;
if(!"") {
  falsyValue = true;
}
falsyValue // Answers true

falsyValue = false;
if(0) {
  falsyValue = true;
}
falsyValue // Answers false
\end{lstlisting}

\paragraph{Equality coercion.} When two objects are compared (as via
the equality operator \ct{==}), depending on their types, none, one or
both objects are coerced before being compared. After coercion, if
both operands have the same type, the comparison is finally resolved
by the strict equality operator. Depending on the type, \ct{valueOf()}
or \ct{toString()} may be implicitely performed on the compared
objects. In addition, when performing an equality comparison, the
following objects are considered as \ct{false}: \ct{null}, \ct{0},
\ct{false}, \ct{''} and \ct{NaN}.

\begin{lstlisting}[caption={Some unintuitive examples of type coercion}, label={lst:type_coercion}]
false == 0 // answers true

0 == false // answers true

"" == 0 // answers true

false == "" // answers true

{} == {} // answers false

var n = {
    valueOf: function () {
        return 1;
    },
    toString: function () {
        return "2";
    }
};

n == 1; // answers true
n == "2"; // answers false

var n = {
    toString: function () {
        return "2";
    }
};

n == 1; // answers false
n == "2"; // answers true

[ [ [ 42 ] ] ] == 42; // answers true. valueOf() an array with one element answers its element

true + 3; // answers 4
\end{lstlisting}

\paragraph{Strict equality coercion.} The strict equality operator
\ct{===} compares both type and value of the operands, without
performing type coercions. It only resolves to true, when both
operands have the same type and value. The only exception are
non-primitive objects, which are strictly equal if they are the same
object.

\begin{lstlisting}
false === 0 // answers false
0 === false // answers false
"" === 0 // answers false
false === "" // answers false

1 === 1 // answers true
{} === {} // answers false

\end{lstlisting}


\chapter{\js programming practices}

This section presents different coding practices in \js development
(\ecma 3) that result in robust, extensible and understandable
sofware.  In particular, we stress the semantics of the \new
constructor and its impact on inheritance \cite{Crock08a}.

\section{Defining prototypes}
\label{sec:prototypes}

\js as defined in \ecma 3 is a prototype-based object-oriented
language where each object has a prototype (referenced in the
\ct{\proto} core property). The object inherits all properties from
it. Since the prototype is also an object, the prototype chain
inheritance is similar to a class hierarchy in class-based
object-oriented languages.

\begin{figure}
  \centering
  \includegraphics[width=.5\textwidth]{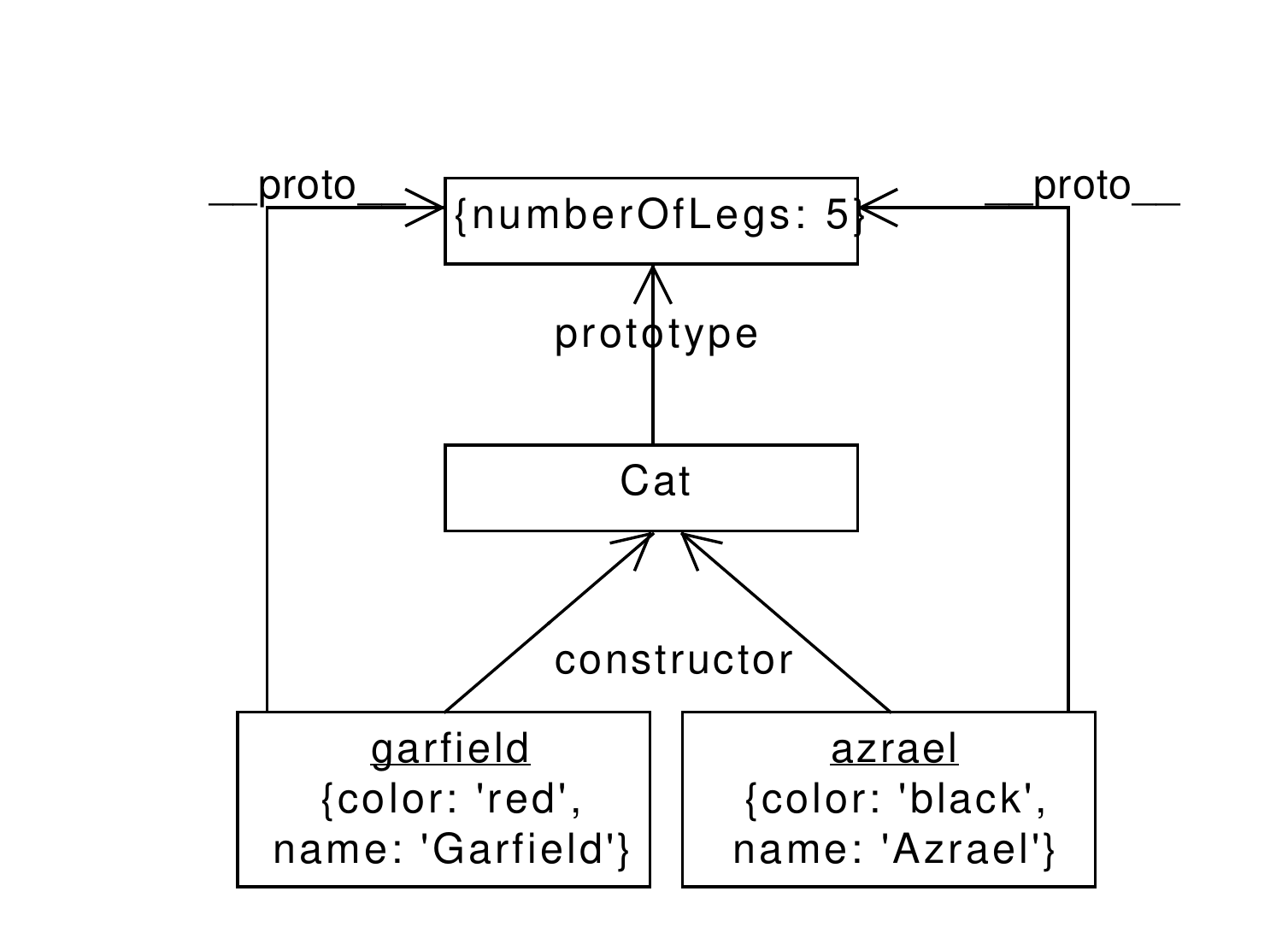}
  \caption{Sharing a property between objects (see \autoref{lst:sharing-through-prototype})}
  \label{fig:sharing-through-prototype}
\end{figure}

Constructors structure object creation and initialize properties. Each
time an object is created, a different copy of each attribute
specified in the constructor is assigned to the object. When sharing
is needed between objects, the shared properties must be defined in
the constructor's prototype as can be seen in
\autoref{lst:sharing-through-prototype} and
\autoref{fig:sharing-through-prototype}.

\begin{lstlisting}[caption={Sharing a value between objects through their prototype (see \autoref{fig:sharing-through-prototype})}, label={lst:sharing-through-prototype}]
var Cat = function (color, name) {
  this.color = color;
  this.name = name || 'default name';
}

Cat.prototype.numberOfLegs = 4;

var garfield = new Cat('red', 'Garfield');
var azrael = new Cat('black', 'Azrael');

garfield.color; // answers 'red'
garfield.numberOfLegs; // answers 4

azrael.color; // answers 'black'
azrael.numberOfLegs; // answers 4

Cat.prototype.numberOfLegs = 5;
garfield.numberOfLegs; // answers 5
azrael.numberOfLegs; // answers 5

azrael.color = 'grey';
garfield.color; // answers 'red'
azrael.color; // answers 'grey'
\end{lstlisting}

\begin{figure}
  \centering
  \includegraphics[width=.5\textwidth]{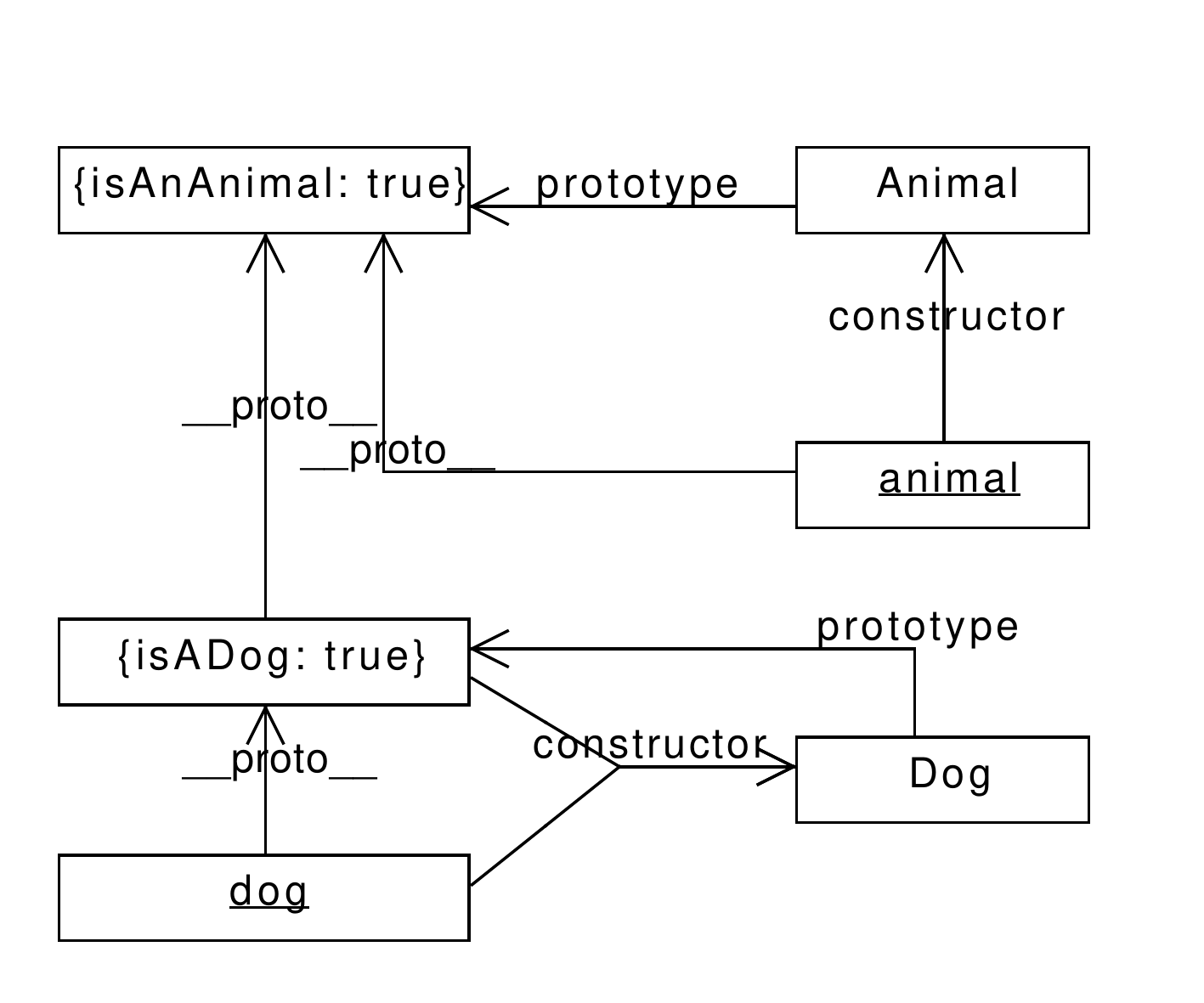}
  \caption{Prototypical inheritance (see \autoref{lst:prototypical-inheritance})}
  \label{fig:protypical-inheritance}
\end{figure}

When a new object is created by calling a constructor, its \ct{\proto}
core property is initialized to the constructor's \ct{prototype}
property value. In \autoref{lst:prototypical-inheritance} we set the
prototype of function \ct{Dog} to be an object created from function
\ct{Animal} (see \autoref{fig:protypical-inheritance}). Note that we
have to set the constructor of the prototype to be the function
\ct{Dog}. Then when an object created from the function \ct{Dog} is
asked for a property that it does not define locally, the lookup is
performed following the prototype chain (\ie, looking inside the
\ct{\proto} core property value). Here the property \ct{isAnAnimal} is
found in the prototype of \ct{Dog} which is an object created from the
\ct{Animal} constructor.

\begin{lstlisting}[caption={Inheritance in \js prototypical programming (see \autoref{fig:protypical-inheritance})}, label={lst:prototypical-inheritance}]
var Animal = function () { };

Animal.prototype.isAnAnimal = true;

var animal = new Animal();

var Dog = function () {};

// The prototype of Dog is set to a new Animal,
// so that future Dog objects will inherit from Animal.prototype
Dog.prototype = new Animal();

// We need to manually change Dog.prototype.constructor so that
// future Dog objects will have Dog as constructor
// (instead of Animal).
Dog.prototype.constructor = Dog;

// All Dog objects must share this property
Dog.prototype.isADog = true;

var dog = new Dog();
dog.isAnAnimal; // answers true
dog.isADog; // answers true
\end{lstlisting}

\begin{figure}
  \centering
  \includegraphics[width=.5\textwidth]{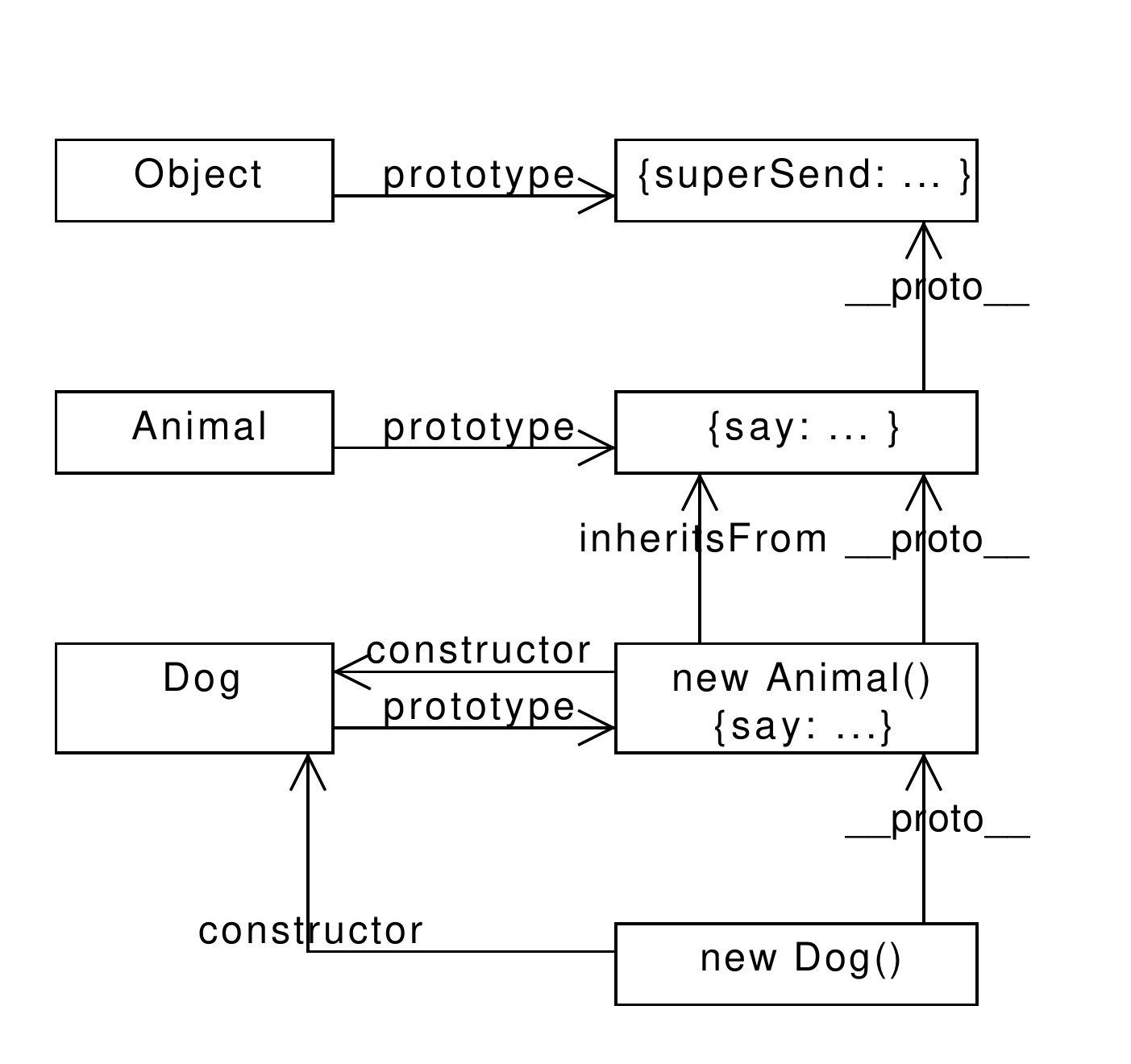}
  \caption{Message resending -- super sends (see \autoref{lst:supersends})}
  \label{fig:supersends}
\end{figure}

\paragraph{Accessing overridden functions.} Other object-oriented
languages have a message resend mechanism, often implemented as
\emph{super} sends. To perform \emph{super} sends in \js, the lookup
must be explicitly forwarded to the prototype (see
\autoref{lst:supersends} and \autoref{fig:supersends}).

\begin{lstlisting}[caption={Message resending -- super sends (see \autoref{fig:supersends})},label={lst:supersends}]
// 'Object' being a function, we add a new method to all objects
Object.prototype.superSend = function (selector, args) {
  // We use 'inheritsFrom' to reference the prototype and we search
  // the property in variable 'selector' from this prototype:
  return this.inheritsFrom[selector].apply(this, args);
};

var Animal = function () { };
Animal.prototype.say = function (string) {
    return 'hello ' + string;
};

var Dog = function () { };
Dog.prototype = new Animal();
// We add our own property to retain inheritance
// without using the not standard __proto__
Dog.prototype.inheritsFrom = Dog.prototype.constructor.prototype;
Dog.prototype.constructor = Dog;

new Dog().inheritsFrom === new Dog().__proto__.__proto__; // answers true

Dog.prototype.say = function (string) {
    return this.superSend('say',['wouf wouf']);
};

new Dog().say("I'm a dog"); // answers 'hello wouf wouf'
\end{lstlisting}

The dynamic capabilities of \js allow the usage of this mechanism to
extend existing objects like \emph{Arrays}, \emph{Numbers}, and
\emph{Functions}, through the addition of properties and methods to
their prototype constructors.

\section{Closures and `functional' inheritance}

We've previously shown an example of how to provide an
inheritance-like relation in \js using prototypes, allowing us to
share properties and methods between our objects. Unfortunately, the
builtin inheritance mechanism of \ecma 3 has several drawbacks: (1) it
depends on many implementation details that could lead to several
mistakes, (2) it does not provide any access protection between
objects, their prototypes, and the outer scope.

In \autoref{lst:closureConstructor} a \js idiom appears showing how we
can use closures to support visibility principles such as private
attributes and inheritance \cite{Crock08a} instead of the typical
prototype relationship we just described. The main idea is to create
functions that create and initialize an object -- declared as
\ct{that} in the example -- with default values and methods. The
values set in the object's properties will be public values. We can
also define and use variables and functions with the \ct{var} keyword
to make them private to our object. To customize the initialization,
we can also pass an object as a parameter to this function as a spec
hash-table.

\begin{lstlisting}[caption=Using closures to support access visibility to properties, label={lst:closureConstructor}]
var animal = function (spec) {
  // We take either the parameter or the empty object if
  // spec is null
  spec = spec || {};
  var that = {};

  // Initialization
  that.isAnAnimal = true;

  // Private
  var name = spec.name || 'unnamed';

  // Public
  that.getName = function() {
    return name;
  };

  return that;
};

var dog = function (spec) {
  spec = spec || {};
  var that = animal(spec); // makes dog inherits from animal

  that.isADog = true;

  return that;
};

var aDog = dog({name: 'milou'});
aDog.isAnAnimal; // answers 'true'
aDog.isADog;     // answers 'true'
aDog.getName();  // answers 'milou'
aDog.name;       // answers 'undefined'
\end{lstlisting}

\chapter{\ecma 5}

We also present an overview of the features of \ecma 5 \cite{ECMA97a, ECMA11a}.

Released in June 2011, \ecma 5 defines the latest standarized version
of the \js language. This release includes improvements and
clarifications compared to \ecma 4. In this section we propose a
survey of two important aspects of \ecma 5: object creation and
properties access. These aspects improve object encapsulation, giving
a finer-grained object-property definition and thus improving
security.

\section{Object creation}

\ecma 5 offers a new way to instantiate objects with the ability to
specify the new object's prototype and properties.

\begin{lstlisting}[caption={Creating an object with null as prototype}, label={lst:create_object_null_ecma5}]
var objOld = new Object();
var objNew = Object.create(null); // new function from ECMAScript 5

// a.isPrototypeOf(b) checks if 'a' is in the __proto__ inheritance
// chain of b (i.e., b is derived from a)
Object.prototype.isPrototypeOf(objOld);  // answers true
Object.prototype.isPrototypeOf(objNew);  // answers false

objOld.toString;  // answers a function
objNew.toString;  // answers 'undefined'
\end{lstlisting}

Example \ref{lst:create_object_null_ecma5} shows how to create a new
object named \ct{objNew} that has no prototype and no inherited
property.

When passed an argument, \ct{Object.create} sets the \ct{\proto}
property of the object to the argument. As a result, \ct{new Object()}
is equivalent to \ct{Object.create(Object.prototype)}.

\ct{Object.create} also optionally accept a second argument that is an
object whose properties serve to initialize the new object. In
\autoref{lst:create_object_ecma5}, the object \ct{obj} inherits all
standard properties from \ct{Object.prototype} and defines a new
property \ct{foo} whose value is the string \ct{"hello"}.

\begin{lstlisting}[caption=Creating an object with prototype and a property, label={lst:create_object_ecma5}]
var obj = Object.create(Object.prototype, {
  foo: { writable: true, configurable: true, value: "hello" },
});
obj.__proto__ === Object.prototype; // answers true
obj.toString; // answers a function
obj.foo; // answers "hello"
\end{lstlisting}

The keys \ct{writable} and \ct{configurable} are described below.

\section{Defining object properties}

Security wise, \ecma 3 doesn't have any concept of private
properties.\footnote{We call private properties object properties that
  are neither enumerable nor writable.} All object properties are
publicly visible, inherited and modifiable at will. \ecma 5 improves
the situation by introducing a fine-grain protocol to define object
properties.

\paragraph{\ct{Object.defineProperty}} is one of the core changes to
\js defined in \ecma 5. This function takes three parameters: the
object on which the property is defined, a string (name of the new
property) and a descriptor object. The descriptor object can be a data
descriptor or a getter/setter descriptor. A descriptor object can have
the following optional keys:

\begin{itemize}
\item \ct{enumerable}: if \ct{true}, the property shows up during
  enumeration of the properties of the receiver;
\item \ct{configurable}: if \ct{true}, the property can be deleted,
  and the descriptor can be changed afterwards.
\end{itemize}

In case of an accessor descriptor, two keys \ct{get} and \ct{set} can
be used to define accessor methods to the property.

In case of a data descriptor, two keys \ct{value} and \ct{writable}
can be used to respectively set an initial value and specify if the
property can be written.

\begin{lstlisting}[caption=Defining properties, label={lst:defineProperty}]
var dog = {};
Object.defineProperty(dog, 'name', {
    enumerable: true,
    configurable: false,
    value: 'Pilou',
    writable: false
});

dog.name; // answers 'Pilou', the default value
dog.name = 'another name'; // tries to set a new value
dog.name; // answers 'Pilou' as the property is not writable

delete dog.name; // tries to remove the property from the object
dog.name; // answers 'Pilou' as the property is not configurable

\end{lstlisting}

Example \ref{lst:defineProperty} shows how to use
\ct{Object.defineProperty}. First an empty object \ct{dog} is
created. A property \ct{'name'} is added to \ct{dog} and set a default
value of \ct{'Pilou'}. This property is neither configurable nor
writable. As a result, trying to change the value of the property
\ct{dog.name} or trying to delete it both fail.

\paragraph{\ct{Object.preventExtensions}.} \ecma 5 introduces two
important functions regarding object extensions:
\ct{preventExtensions} and \ct{isExtensible}, both available from
\ct{Object}. As seen in \autoref{lst:preventExtensions}
\ct{preventExtensions} takes an object as argument and prevents any
further addition of properties. Existing properties may still be
deleted though. \ct{isExtensible} is a testing function answering a
boolean value that indicates whether properties can be added or not.

\begin{lstlisting}[caption=Preventing object extensions, label={lst:preventExtensions}]
var dog = {};
Object.defineProperty(dog, 'name', {
    enumerable: true,
    configurable: false,
    value: 'Pilou',
    writable: false
});

Object.isExtensible(dog); // answers true
Object.preventExtensions(dog);
Object.isExtensible(dog); // answers false

dog.age = 5; // tries to add a new property to 'dog'
dog.age // anwers undefined because 'dog' is not extensible
\end{lstlisting}

\ecma 5 also introduces full immutability of objects through \ct{Object.freeze}, and can be tested with \ct{Object.isFrozen}, and seen in \autoref{lst:freeze}.

\begin{lstlisting}[caption=Object immutability, label={lst:freeze}]
var dog = {};
Object.defineProperty(dog, 'name', {
    enumerable: true,
    configurable: false,
    value: 'Pilou',
    writable: false
});

Object.isFrozen(dog); // answers false
Object.freeze(dog);
Object.isFrozen(dog); // answers true

dog.age = 5; // tries to add a new property to 'dog'
dog.age // anwers undefined because 'dog' is not extensible

delete dog.name // answers false
dog.name // answers 'Pilou'
\end{lstlisting}

By adding the functions mentioned in this section (notably
\ct{create}, \ct{defineProperty}, \ct{preventExtensions} and \ct{freeze}), \ecma 5
makes it possible for developers to secure their objects.

\chapter{Key Bibliography Elements}

Here we present the key and limited articles that we encourage you to
read if you want to get deeper into the semantics of \js.

\begin{itemize}
\item Sergio Maffeis, John C. Mitchell, and Ankur Taly. An operational
  semantics for javascript. In Proceedings of the 6th Asian Symposium
  on Programming Languages and Systems, APLAS '08, pages 307-325,
  Berlin, Heidelberg, 2008. Springer-Verlag. The paper presents a
  30-page operational semantics, based directly on the JavaScript
  specification. The semantics covers most of JavaScript directly, but
  does omit a few syntactic forms. They discuss various differences
  between the standard and implementations.

\item Arjun Guha, Claudiu Saftoiu, and Shriram Krishnamurthi. The
  essence of javascript. In Proceedings of the 24th European
  Conference on Object-Oriented Programming, ECOOP'10, pages 126-150,
  Berlin, Heidelberg, 2010. Springer-Verlag. In this article a core
  calculus is defined. Based on it, several aspects of JavaScript are
  described.  Some badly designed features of Javascript are
  described.

\item ECMAScript version 3.0 specification (188
  pages). \url{http://www.ecma-international.org/publications/files/ECMA-ST-ARCH/ECMA-262,\%203rd\%20edition,\%20December\%201999.pdf}
\item ECMAScript version 5.1 specification (255
  pages). \url{http://www.ecma-international.org/publications/files/ECMA-ST-ARCH/ECMA-262\%205th\%20edition\%20December\%202009.pdf}
\end{itemize}

\chapter{Conclusion}

This deliverable has introduced an overall picture of \js as defined
in \ecma 3 and \ecma 5, focusing on security aspects of its semantics.
We offered a detailed review of \js subtle points concerning security
and application predictability.

We have seen that the \window object is key to data privacy
protection. Nevertheless, we have pointed out how \js exposes
\window to the entire environment constraining any security
approach.

We have described scoping issues and the possible resulting security
leaks. For example, the \this pseudo variable is bound to a different
object depending on the way the function invocation expression is
written (syntactic shape of the function invocation in which \this is
used). This dynamic behavior can be exploited by an attacker to leak
private objects. As another example, the fact that \js lifts variable
definitions inside functions (\ie moves to the top) leads to
unsuspected variable shadowing, dramatically lowering behavior
predictability.

In the last chapter, we have proposed a description of \ecma 5
features regarding object declaration and property access that improve
encapsulation, therefore securing exposure of objects.

In the following deliverable, we will leverage this knowledge,
detailing existing sandboxing techniques, and for each of them, we
will review its advantages and weaknesses in different contexts.



\bibliographystyle{alpha}

\end{document}